\documentclass[12pt]{article}

\usepackage{graphicx}
\usepackage{amssymb}

\begin{document}

\title{Spectral representation of the Casimir Force Between a Sphere and a Substrate} 

\author{C. E. Rom\'an-Vel\'azquez \\
\it {\small Centro de Investigaci\'on en Ciencia Aplicada y Tecnolog\'{\i}a Avanzada,}
\\ \it {\small Instituto Polit\'ecnico Nacional, Av. Legaria 694, Col. Irrigaci\'on, D.F. 11500, M\'exico}\\
\rm Cecilia Noguez, C. Villarreal, and R. Esquivel-Sirvent\\
\it {\small Instituto de F\'{\i}sica, Universidad Nacional Aut\'onoma de M\'exico,}\\ 
\it {\small Apartado Postal 20-364, D.F. 01000,  M\'exico}}

\date{\small Received: December 18, 2002; Revised: February 18, 2003}
\maketitle

\begin{abstract}
We develop a spectral representation formalism to calculate the Casimir force, in the non-retarded limit, or van der Waals force between a spherical particle and a substrate, both with arbitrary local dielectric properties. The spectral formalism allows to study the system as a function of its geometrical properties separately from its dielectric properties. 
The calculated force is attractive, and at a small separations it is orders of magnitude larger for nanometric-size spheres than for micrometer particles. We also found that the force depends more on the dielectric properties of the sphere  than  of the substrate. 
\end{abstract}


Recent advances in micro and nano devices have opened the possibility of studying quantum phenomena that occur at these length scales. Such is the case of the Casimir force~\cite{casimir} that is a macroscopic manifestation of the quantum vacuum fluctuations, as predicted by quantum electrodynamics. The textbook example~\cite{milonni,milton,ninham} consists of two parallel neutral conducting plates which attract each other. The first experimental measurements were done in 1951 using dielectric materials~\cite{primer}, and in 1958 using conductors~\cite{sparnay}. These measurements have large errors, and up to recently, it was possible to perform measurements with about 15\% of precision on truly parallel metal surfaces~\cite{bressi}.
The difficulty of keeping the two plates parallel at separations of few nanometers  makes it easier to measure the Casimir force between a sphere and a plane~\cite{phgm,lamoraux,chan,mohideen,mohideen2}. In this case, the Casimir theory for parallel plates can be extended using the proximity theorem~\cite{primer}. 
The approximation is valid when the minimum separation between the sphere and the plane is much smaller than the radius of the sphere. This theorem was employed to corroborate experimental measurements of the Casimir force between a plane and a large sphere~\cite{chan,mohideen,mohideen2}. However, it is well known that quantum effects become more evident as the size of the system decreases. Thus, the question of how important are the Casimir effects on nanometric-size spheres is still an open question of fundamental importance.  
In 1948 Casimir and Polder~\cite{casypol} calculated the force of a polarizable atom near a perfect conductor plane considering the influence of retardation, and finding a correction to the London or van der Waals forces. Retardation effects are important if we consider that the distance between the atom and the plane is larger than the characteristic length of the system. Complementary theories are necessary to handle nanometer-size systems with real dielectric properties. Within this context Ford~\cite{ford} calculated the force between a perfectly conducting wall and a sphere with a Drude dielectric function. After a delicate cancelation of terms in the equations, he obtained a force that changes from attractive to repulsive in an oscillatory fashion depending on the relative distance between the sphere and the surface. However, this oscillatory behavior has not been observed experimentally~\cite{phgm,lamoraux,chan,mohideen,mohideen2}, and has not been predicted by other theories~\cite{primer,casypol}.

In this work, we develop a spectral representation formalism to calculate the force between a sphere and a substrate. The advantage of this spectral representation is that we can separate the contribution of the dielectric properties of the sphere and substrate from the contribution  of its geometrical properties. Since results for large spheres~\cite{primer} and large distances~\cite{casypol} are known, we restrict ourselves to the case of nanometric-size spheres and distances of few nanometers. In this case, it is not necessary to consider retardation effects, therefore, we work in the quasi-static limit such that  the radius of the sphere and the minimum separation between the sphere and the plane, are smaller than the characteristic length of the system~\cite{longitud}. In this regime, the Casimir force is commonly known as the van der Waals or London force~\cite{ninham}.

 \begin{figure}[tbh]
\centerline {\includegraphics[width=3in]{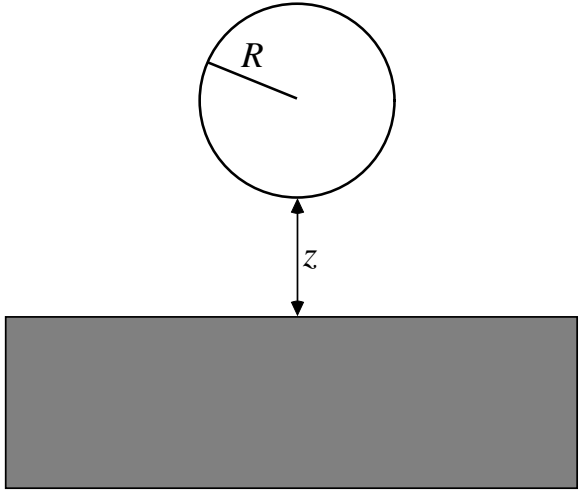}} 
\caption{Schematic model of the system.}
 \end{figure}

We consider a homogeneous sphere of radius $R$, electrically neutral and with a local dielectric function $\epsilon_{\rm sph}(\omega)$. The sphere is suspended at a minimum distance $z$ above a substrate (see Fig.~1) which is also neutral and has a local dielectric function $\epsilon_{\rm sub}(\omega)$. The space or ambient between the sphere and substrate is  vacuum ($\epsilon_{\rm amb} = 1$). The quantum fluctuations of the electromagnetic field induce a polarization in the sphere which can be described by a point dipole located at its center,
\begin{equation}
{\mathbf p}_{\rm sph} (\omega) = \alpha(\omega) {\bf E}^{\rm vac} (\omega), \end{equation}
where $\alpha(\omega) = [\epsilon_{\rm sph}(\omega) - \epsilon_{\rm amb} ] / [\epsilon_{\rm sph}(\omega) + 2\epsilon_{\rm amb}] R^3 $, is the polarizability of the sphere that is assumed to be polarized uniformly~\cite{alfa}, and ${\bf E}^{\rm vac} (\omega)$ is the electromagnetic field associated to the vacuum fluctuations. We can rewrite the polarizability as
\begin{equation}
\alpha(\omega) = \frac{n_0} {n_0 - u (\omega)} R^3, \label{polariza}
\end{equation}
where $n_0 = 1/3$ is a constant and  $u(\omega) = [1 - \epsilon_{\rm sph}(\omega)/ \epsilon_{\rm amb}]^{-1}$ is a variable that only depends on the dielectric properties of the sphere and the ambient. When the sphere is near a substrate, it induces a charge distribution on the substrate that can be seen as a dipole image, such that
\begin{equation}
{\bf p}_{\rm sub}(\omega) = f_c(\omega) {\mathbb M} \cdot {\bf p}_{\rm sph}(\omega),
\end{equation}
where the term $f_c{\mathbb M}$ satisfies the boundary conditions of the system, being ${\mathbb M}$ a diagonal matrix whose elements depend on the choice of the coordinate system, and $f_c(\omega) = [\epsilon_{\rm amb} - \epsilon_{\rm sub}(\omega)] / [\epsilon_{\rm amb} + \epsilon_{\rm sub}(\omega)] $ a the contrast factor that only depends on the dielectric properties of the substrate and the ambient. The induced charge distribution on the substrate produces a field which also modifies the sphere's dipole moment through a local field. Thus, the total induced dipole-moment on the sphere is  
\begin{equation}
{\bf p}_{\rm sph}(\omega) =  \alpha(\omega) \left[ {\bf E}^{\rm vac}(\omega) + {\mathbb T} \cdot {\bf p}_{\rm sub}(\omega) \right], \label{dipolo}
\end{equation}
where ${\mathbb T}$ is the dipole-dipole interaction tensor and, in the non-retarded limit, it takes the form 
\begin{equation}
{\mathbb T} = \frac{3 \hat{\bf r} \hat{\bf r} - {\mathbf 1}}{r^3},
\end{equation}
where ${\mathbf 1}$ is a unitary matrix, ${\bf r}= (0,0,2(z+R))$ is the vector from the center of the image dipole to the center of the sphere, $\hat{\bf r} = {\bf r}/ r$, and $r = |{\bf r}|$. Given the symmetry of the system the diagonal components of ${\mathbb M}$ are ($-1,-1,1$), and there are only three independent components of ${\mathbb T}$, one perpendicular to the surface plane and two parallel to this plane.  

The frequencies that satisfy the boundary conditions of the system are those at which the sphere is polarized. These frequencies are known as the proper electromagnetic modes of the system, and we denote them like $\omega_s$. Then, the total energy of the system is  ${\cal E} = \sum_s 1/2 \hbar \omega_s$.  A convinient way of determining these  proper electromagnetic modes is using a spectral representation formalism that we derive as follows.

First we rewrite Eq.~(\ref{dipolo}) using the expression of the polarizability from Eq.~(\ref{polariza}), as
\begin{equation} 
\left[ - u(\omega) {\mathbf 1} + {\mathbb H} \right] \cdot {\bf p}_{\rm sph}(\omega) = {\mathbf V}^{\rm vac}, \label{matriz}
\end{equation}
where ${\mathbf V}^{\rm vac} = n_0 R^3{\mathbf E}^{\rm vac}$, and ${\mathbb H} = n_0 [{\mathbf 1} - f_c(\omega) R^3 {\mathbb T}{\mathbb M}]$  is a dimensionless matrix that only depends on the geometry of the system. To find the solution of  Eq.~(\ref{matriz}), consider the case when $f_c(\omega)$ is real, then ${\mathbb H}$ is a real and symmetric matrix. In this case, we can always find a unitary transformation that diagonalizes it,  ${\mathbb U}^{-1}{\mathbb H}{\mathbb U} = n_s$, being $n_s$ the eigenvalues of ${\mathbb H}$. Furthermore, the solution of Eq.~(\ref{matriz}) is given by
\begin{equation}
{\bf p}_{\rm sph}(\omega) = {\mathbb G}(u) {\mathbf V}^{\rm vac},
\end{equation}
where ${\mathbb G}(u) = [ - u(\omega) {\mathbf 1} + {\mathbb H}]^{-1}$ is a Green's operator. The $ij$th element of ${\mathbb G}(u)$ can be written in terms of the unitary matrix ${\mathbb U}$ as~\cite{libro}
\begin{equation}
G_{ij}(u) = \sum_s \frac{U_{is}(U_{js})^{-1}}{u-n_s}. \label{green}
\end{equation}
The poles of ${\mathbb G}(u)$, that is $u(\omega) = n_s$, give the frequencies of the proper electromagnetic modes, $\omega_s$, of the system~\cite{new}. We now calculate the Casimir interaction energy as the difference between the energy when the sphere is at a distance $z$ from the substrate and the energy when $z \to \infty$, that is,
\begin{equation}
{\cal E} = \sum_s \frac{\hbar \omega_s}{2} - \sum_{s'} \frac{\hbar \omega_{s'}}{2}. \label{ener}
\end{equation}
The eigenfrequencies $\omega_{s'}$ are obtained from the poles of Eq.~(\ref{green}) when $z \to \infty$, or by substituting  $\epsilon_{\rm sub}(\omega) = \epsilon_{\rm amb}$ in $f_c$. Note that it is not necessary to do any renormalization or any delicate cancelation to calculate the energy. Alternatively, we can also find the density of states using the Green's function definition and then calculate the energy of the system, as we show in the appendix.

The advantage of the spectral representation is that we can separate the contribution of the dielectric properties of the sphere from the contribution of its geometrical properties. As we mentioned, the material properties of the sphere are contained in the spectral variable $u$, while the geometrical properties of the system, like the radius of the sphere and the separation of the sphere to the substrate are in the matrix ${\mathbb H}$. Furthermore, ${\mathbb H}$ is a dimensionless matrix that depends on the ratio $z/R$. Its eigenvalues are independent of ${\mathbf V}^{\rm vac}$ and of the dielectric properties of the sphere. And the dielectric properties of the substrate are in $f_c$ which is a real function even for dispersive materials~\cite{rota}. A similar spectral representation was proposed years ago to study the effective dielectric properties of granular composites~\cite{fuchs}.
The results discuss here are calculated as follows. First, we construct the matrix $\mathbb H$ for a given $z/R$, and we diagonalize it numerically to find its eigenvalues $n_s$. Considering an explicit expression for the dielectric function of the sphere, we calculate the proper electromagnetic modes $\omega_s$ trough the relation $u(\omega_s) = n_s$. Once we have $\omega_s$, we calculate the energy according with Eq.~(\ref{ener}). Here, we use the Drude model, such that $\epsilon_{\rm sph}(\omega) = 1 - \omega_p^2/[\omega(\omega + i/\tau)]$, where $\omega_p$ is the plasma frequency and $\tau$ is the relaxation time.
We present results for potassium (K), gold (Au), silver (Ag) and aluminum (Al) spheres  with $\hbar \omega_p = 3.80$, 8.55, 9.60, and 15.80~eV, and $(\tau \omega_p)^{-1} = 0.105$, 0.0126, 0.00188, and 0.04, respectively.  We have considered substrates whose dielectric function is real and constant in a wide range of the electromagnetic spectrum as sapphire (Al$_3$O$_2$), and titanium dioxide (TiO$_2$), with $\epsilon_{\rm sub} = 3.13$, and 7.81, respectively. Then, the corresponding contrast factors are $f_c=$ -0.516, and -0.773. We have also considered the case of a perfect conductor substrate (denoted by Inf) with $\epsilon_{\rm sub}  \to \infty$ and $f_c=-1$. 

\begin{figure}[tbh]
\centerline {
\includegraphics[width=5.2in]{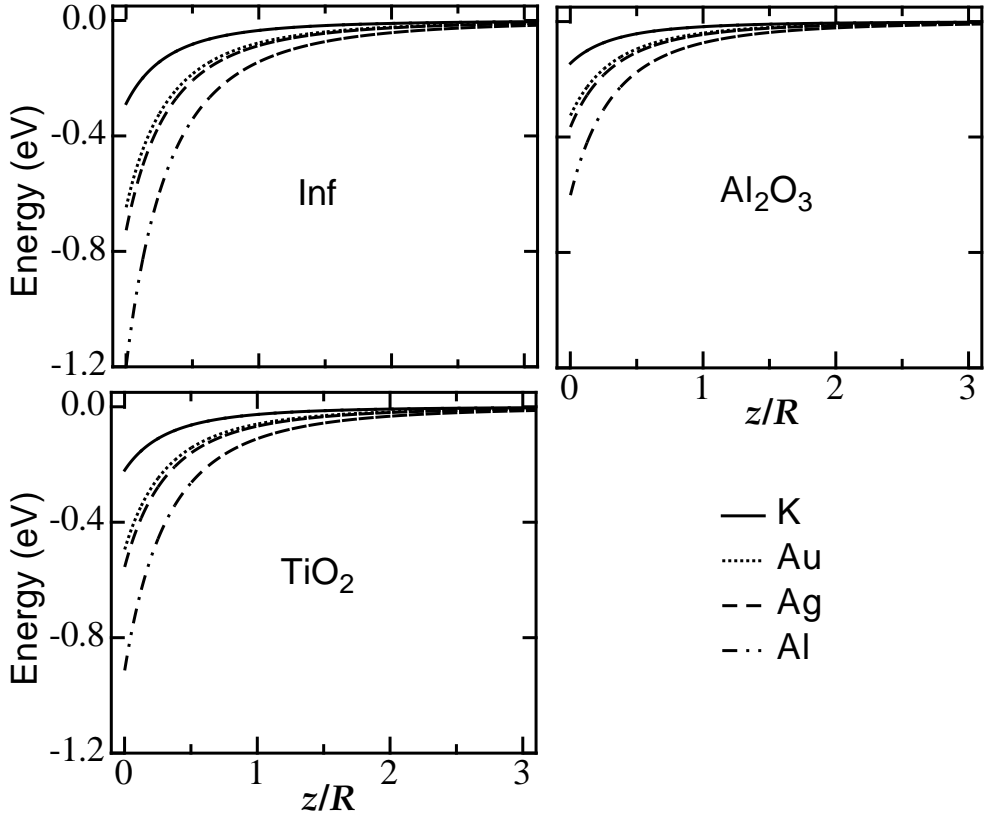}
} 
\caption{Energy as a function of $z/R$. Each panel shows the results for different substrates (perfect conductor, Al$_3$O$_2$, and TiO$_2$).}
\end{figure}

In Fig.~2, we show the energy as a function of $z/R$. In general, we observe that the energy shows a power law of $(z/R)^{-3}$. This behavior is independent of the material properties, and it is inherent to the dipole-dipole interaction model. This is consistent with the result found by Casimir and Polder~\cite{casypol} for a polarizable atom, and with the measurements by Mohideen {\it et al}.~\cite{mohideen}, but it is contrary to the oscillatory behavior calculated by Ford~\cite{ford}. The value of the energy varies with the substrate, for example, at small distances it is about two times larger for a perfect conductor substrate than for Al$_3$O$_2$, while the TiO$_2$ case is between them. This is easily explained if we look at the contrast factor values for each substrate, where one can see that as $f_c \to -1$, the energy is larger. For all the substrates, we found that $V$ becomes larger as the plasma frequency of the metal also does. In conclusion, we found that $V$ is large when $f_c \to -1$ and $\omega_p$ is  large, recovering the limit for perfect conductors. Therefore, the energy is largest (smallest) for an Al (K) particle over a perfect conductor (Al$_3$O$_2$) substrate. When the sphere is at a distance larger than $2R$, the energy is very similar, independently of the dielectric properties of the sphere and substrate.

\begin{figure}[tbh]
\centerline {
\includegraphics[width=5.45in]{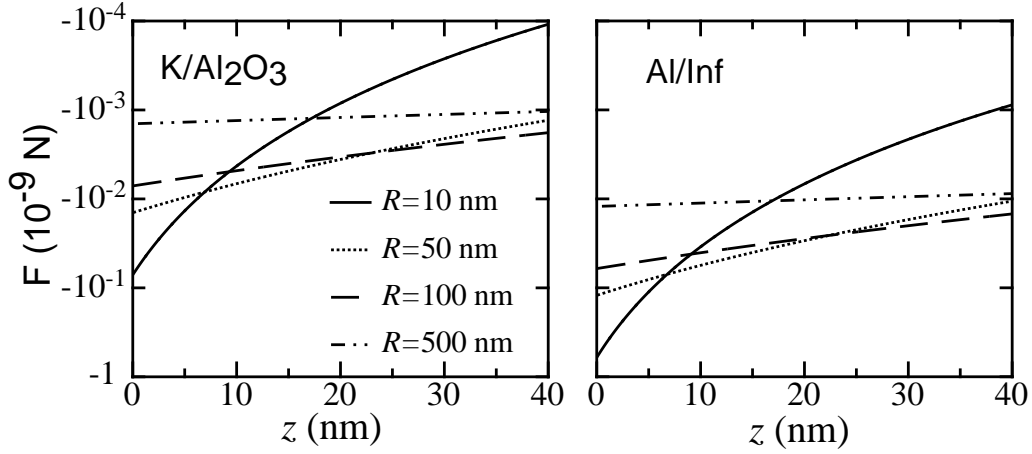}
} 
\caption{Casimir force as a function of $z$. In the left (right) panel we show results for a K (Al) sphere over an Al$_3$O$_2$ (Inf) substrate.  The different curves correspond to spheres of 10~nm, 50~nm, 100~nm and 500~nm of radius.}
\end{figure}

Let us first analyze the force as a function of the geometrical properties, that is, as a function of $R$ and $z$. In Fig.~3 we show the Casimir force calculated as 
\begin{equation}
F = - \frac{d{\cal E}}{dz} = -\frac{1}{R} \frac{d{\cal E}}{d(z/R)}.
\end{equation} 
In all cases, we obtain an attractive force such that, as $R$ is smaller the force increases. When the sphere is almost touching the substrate ($z\sim0$~nm) the force is ten times larger for a sphere of $R=10$~nm than the one of $R=100$~nm, and increases fifty times for a sphere of $R=500$~nm. As a function of $z$, the force for the sphere with $R=500$~nm seems to be almost constant from 0 to 40~nm compared with the other curves. This is an artifact of the scale since all curves have a power law of $z^{-4}$, and they are proportional to $R^{2}$. This implies that for a distance $z \le  10$~nm the force is larger for the smallest sphere; however, at a larger distance the force is larger for larger spheres, while for very large distances, the force is independent of $R$. Furthermore, the force for the sphere with $R=10$~nm decreases about three orders of magnitude as the separation of the sphere goes from 0 to 40~nm, independently of the dielectric properties of the system. On the other hand, with the proper combination of dielectric functions of the sphere and substrate it is possible to modulate the magnitude of the Casimir force. Here, we show the force for an Al sphere over a perfect conductor which is one order of magnitude larger than the force between the K sphere over Al$_3$O$_2$, as it is expected.  

\begin{figure}[tbh]
\centerline {
\includegraphics[width=5.45in]{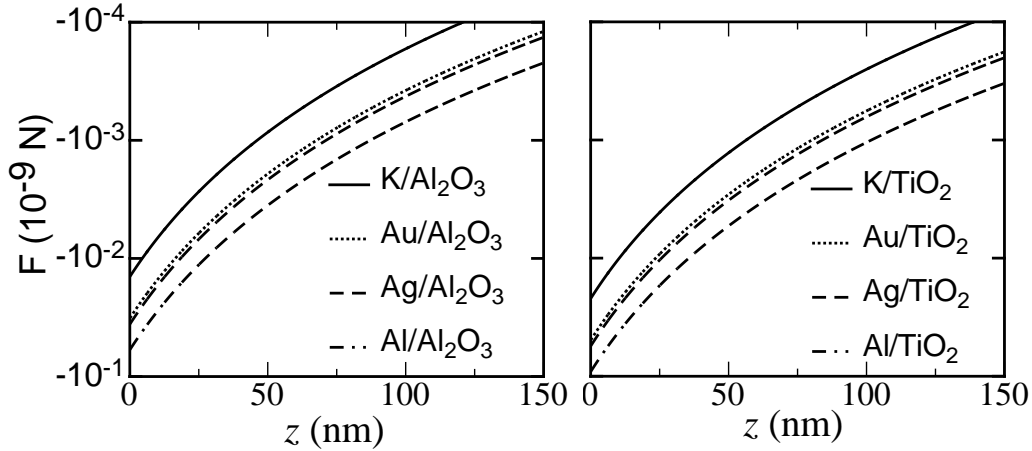}
} 
\caption{Casimir force as a function of $z$ for spheres of K, Au, Ag, and Al with $R=50$~nm over Al$_3$O$_2$, and TiO$_2$.}
\end{figure}

From Fig.~4 we analyze the force as a function of the dielectric properties of the particles and substrate. In all cases, we found the same dependence of the force as a function of $z$, independently of the dielectric functions of both, sphere and substrate. Although the dielectric function of the substrate is important, the dependence on the dielectric function of the sphere is more critical in the magnitude of the Casimir force. Indeed, the force is larger for increasing values of $\omega_p$. In particular, we found that the force for an Al sphere is almost ten times larger than for a K sphere. On the other hand, for a given sphere the Casimir force increases at most by a factor of three, when the substrate is changed from Al$_3$O$_2$ to a perfect conductor. 

The spectral representation formalism allows us to calculate the force between a sphere and a substrate in a range of sizes and separations where the proximity theorem is not applicable.  For example, for a Au sphere of radius 100 nm on top of perfect conducting plane,  the value of the Casimir force using the proximity theorem yields a force about three orders of magnitude of the values obtained in this paper. This is due to the linear dependence of the proximity theorem with the radius of the sphere and that the geometrical effects of the sphere are not included. On the other hand, systems where the proximity theorem is used, such as the experiments by the group of Mohideen and collaborators ~\cite{mohideen,mohideen2} do not employ homogeneous metallic spheres, rather coated dielectric spheres, making it difficult to employ the spectral representation formalism.  

In conclusion, we have developed a spectral representation formalism within the van der Waals approximation to calculate the Casimir force between a sphere and a substrate. This spectral formalism  separates the geometrical properties contributions from dielectric properties contributions on the Casimir effect in the non-retarded limit.  We found that at very small distances, the force can increase  orders of magnitude as the size of the particle becomes smaller. We have also observed that the correct choice of the dielectric properties of both, sphere and substrate, can increase or decrease the force by orders of magnitude. 

This work has been partly financed by CONACyT grant No.~36651-E and by DGAPA-UNAM grants No.~IN104201 and IN107500.

\appendix
\section{Appendix}

By using the Green's function from Eq.~(8), one can find the density of states of the system as a function of the spectral variable $u$, as $ \rho(u) = -\frac{1}{\pi}  {\rm Im}\left[ {\rm Tr}  {\mathbb G}(u) \right] .$ In the particular case, when the sphere is metallic and its dielectric function is described by the Drude model, $u (\omega) = \omega (\omega + i \tau)/\omega_p^2$. Then, the density of states as a function of the frequency takes the following explicit form
\begin{equation}
\rho(\omega) = \frac{2 w_p}{\pi } \sum_s \sqrt{n_s} \left( \frac{ \omega/\tau}{(\omega^2 - \omega_p^2 n_s )^2 + (\omega/\tau)^2} \right), 
\end{equation}
 and the energy can be calculated as $
V = \frac{1} {2}\int_0^\infty \hbar \omega \rho(\omega) d \omega. $




\begin{thebibliography}{99}
\bibitem{longitud} The sphere's radius and the minimum separation between the sphere and the substrate are much smaller than the characteristic length of the system, in this case, by $l=c/\omega_p$, with $c$ the speed of light.
\bibitem{alfa} We have assumed that the sphere is polarized uniformly.  This assumption is valid when the sphere is small, and  is not very close to the substrate.
\bibitem{rota} If the dielectric function of the substrate is described by the Lorentz or Drude formula, we can perform a rotation to the imaginary axis  ($\omega \to i\omega$) to get a real dielectric function.
\end{thebibliography}
\end{document}